\documentstyle[11pt,epsf]{article}
\begin{document}
\begin{titlepage}
\begin{flushright}
IASSNS-93/83\\
WUB-93-40\\
Dec 1993
\end{flushright}
\vspace{4cm}
\begin{center}
\LARGE \bf
Specific Heat Exponent for the 3-d Ising Model\\
from a 24-th Order High Temperature Series.
\normalsize

\vspace{1cm}

{\it by }

\vspace{1cm}

Gyan Bhanot\footnote{Thinking Machines Corporation, 245 First Street, 
                     Cambridge, MA 02142, USA}\footnote{Institute 
                     for Advanced Study, Princeton, NJ 08540, USA},
Michael Creutz\footnote{Brookhaven National Laboratory, Upton, 
                        NY 11973, USA},\\
Uwe Gl\"assner\footnote{Physics Department, University of Wuppertal, 
                        Gaussstrasse 20, 42097 Wuppertal, Germany},
Klaus Schilling$^{4}$

\vspace{1cm} \bf ABSTRACT \\ \end{center} We compute high temperature
expansions of the 3-d Ising model using a recursive transfer-matrix
algorithm and extend the expansion of the free energy to 24th order.
Using ID-Pad\'{e} and ratio methods, we extract the critical exponent
of the specific heat to be $\alpha = 0.104(4)$.  

\end{titlepage} 
\section{INTRODUCTION} 

High- and low-temperature expansions constitute
major tools for the calculation of critical properties in statistical
systems. The Ising and Potts model low temperature
expansions were recently extended 
\cite{creutz,bhanot_ising,bhanot_series,enting} using a technique
based on the method of recursive counting \cite{binder}. In a separate
development, Vohwinkel \cite{vohwinkel} implemented the
shadow-lattice technique of Domb \cite{domb} in a very clever way and
added many new terms to the series.  However, the extraction of
critical parameters from low temperature series is hampered by the
presence of unphysical singularities. This is especially true of the
3-d Ising model.  For this reason, low temperature analytic methods
are very often inferior to Monte-Carlo methods for computing critical
exponents.

High-temperature (HT) expansions on the other hand, generally have
better analytic behavior and yield more accurate exponents.  Very
recently, two variants of the recursive counting technique for HT
expansions have been pursued. While Enting and Guttmann \cite{enting}
keep track of spin configurations on a set of rectangular finite
lattices, ref.\cite{bhanot_series} counts HT-graphs on finite, helical
lattices.  Such computer based series expansions have very large
memory requirements. This makes them ideal candidates for large
parallel computers if communication issues can be handled efficiently.
In this paper we will present the results of a HT expansion of the 3-d
Ising model to 24th order, obtained on a 32 node 1 GByte Connection
Machine CM-5. The implementation is based on a bookkeeping algorithm
of binary coded spin configurations in helical geometry.

\section{COMPUTATION OF THE SERIES}
We start with a discussion of the HT algorithm to compute the partition 
functions on finite 3-d Ising lattices. Starting from the action 
\begin{equation}
E\{s\} = - \sum_{<i,j>} s_{i} s_{j} \mbox{  ,}
\end{equation} 
the partition function is
\begin{equation}
Z = \sum_{ \{s\} } \exp{(-\beta E)} 
  = \sum_{ \{s\} } \prod_{<i,j>} \exp{(\beta s_{i} s_{j})} 
\end{equation}
and is expanded in a HT series \cite{parisi}
\begin{equation}
Z = ( \cosh^3{\beta} )^V \sum_{ \{s\} } \prod_{<i,j>} 
    ( 1 + s_{i} s_{j} t ) 
  = (2 \cosh^3{\beta})^V \sum_k p(k) t^k \mbox{ ,} \label{loop_series}
\end{equation} with the HT expansion parameter $t=\tanh{\beta}$. V is
the volume of the system. The free energy per spin is defined as
\begin{equation} f = -\frac{1}{\beta V} \log{Z} = - \frac{2
\cosh^3{\beta}}{\beta} - \frac{1}{\beta} \sum_k f_{k} t^{k} \mbox{ .}
\end{equation} For simplicity, consider a finite simple cubic lattice
which, in the recursion algorithm, is built up by adding one site
after the other, layer by layer. This procedure defines the recursion
step, which requires knowledge only of those spin states that are
contained in the exposed two-dimensional surface layer. To minimize
finite size effects, it is best to use helical boundary conditions
\cite{bhanot_ising,bhanot_series}. One can visualize helical boundary
conditions by imagining all spins in the layer laid out along a
straight line. In this picture, the nearest neighbours to a given site
in the sequence in the $i$th direction can be chosen to be $h_i$ sites
away, with $i = x,y,z$. It is convenient to assume $h_x < h_y < h_z$.
It is easy to see that as spins are added, one needs only to keep track of 
the states of spins on the topmost $h_z$ sites. Let these spins be denoted
${s_{1},\ldots,s_{h_{z}}}$. Then the partition function can be
rewritten as \begin{equation} Z = (2 \cosh^3{\beta})^V
    \sum_k \sum_{ s_{1},\dots,s_{h_{z}} } 
    p(k;s_{1},\dots,s_{h_{z}}) t^{k} \mbox{  .} \label{series}
\end{equation}
The recursion step, which consists of adding another spin $s_0$ to the 
system, changes the partition function into 
\begin{eqnarray}
\label{link_factors}
Z & = & 2^V (\cosh^3{\beta})^{V+1} 
    \sum_{s_0} \sum_k \sum_{ s_{1},\dots,s_{h_{z}} } 
    p(k;s_{1},\dots,s_{h_{z}}) t^{k} \\ 
  &   &  \times ( 1+s_{0} s_{h_x} t ) 
    ( 1+s_{0} s_{h_y} t ) ( 1+s_{0} s_{h_z} t ) 
\nonumber \mbox{  .}
\end{eqnarray}
$s_{h_x},s_{h_y}$ and $s_{h_z}$ are the backward nearest neighbours of the 
site $s_{0}$. The site $s_0$ will displace its backward $z$ neighbour 
site $s_{h_z}$ after the counting of the added spin is completed. 
Since $s_{h_z}$ will not be referred to in the subsequent steps
of the algorithm, the summation over $s_{h_z}$ can be carried out:
\begin{eqnarray}
Z & = &2^V (\cosh^3{\beta})^{V+1} 
      \sum\limits_{s_{0}} \sum\limits_k 
      \sum\limits_{ s_{1},\ldots,s_{h_z-1} } \nonumber \\
  & \times [ & + p(k;s_{1},\ldots,s_{h_z-1},s_0) t^{k} 
         ( 1+s_{0} s_{h_x} t ) ( 1+s_{0} s_{h_y} t ) ( 1+t )  \\ 
  &  & + p(k;s_{1},\ldots,s_{h_z-1},\bar{s_0}) t^{k} ( 1+s_{0} s_{h_x} t )
        ( 1+s_{0} s_{h_y} t ) ( 1-t ) ] \nonumber \mbox{  .}
\end{eqnarray}
The contribution in the second (third) line of this equation contains 
the part with $s_{h_z}$ being parallel (antiparallel, denoted by $\bar{s_0}$) to
$s_{0}$. Comparing this expression with the HT series (\ref{series}) for 
the new system yields the recursion relation induced for the coefficients 
$p$:
\begin{eqnarray}
   2 p'(k;s_{0},s_{1},\ldots,s_{h_{z}-1}) 
&=& p(k-0;s_{1},\ldots,s_{h_{z}-1},s_0)  \nonumber \\
&+& p(k-0;s_{1},\ldots,s_{h_{z}-1},\bar{s_0}) \nonumber \\
&+& p(k-1;s_{1},\ldots,s_{h_{z}-1},s_0) (s_0 s_{h_x} + s_0 s_{h_y} + 1) 
 \nonumber \\
&+& p(k-1;s_{1},\ldots,s_{h_{z}-1},\bar{s_0}) (s_0 s_{h_x} + s_0 s_{h_y} - 1) 
 \nonumber \\
&+& p(k-2;s_{1},\ldots,s_{h_{z}-1},s_0) 
     (s_{h_x} s_{h_y} + s_0 s_{h_x} + s_0 s_{h_y})  \\
&+& p(k-2;s_{1},\ldots,s_{h_{z}-1},\bar{s_0}) 
     (s_{h_x} s_{h_y} - s_0 s_{h_x} - s_0 s_{h_y}) \nonumber \\
&+& p(k-3;s_{1},\ldots,s_{h_{z}-1},s_0) (s_{h_x} s_{h_y}) \nonumber \\
&+& p(k-3;s_{1},\ldots,s_{h_{z}-1},\bar{s_0}) (- s_{h_x} s_{h_y}) \nonumber 
\end{eqnarray}
It is crucial to remove finite-size errors by combining the results of
different lattice structures as described in refs. 
\cite{bhanot_ising,bhanot_series}. We use the set of lattices 
listed in table \ref{lattices}
\begin{table}
\footnotesize
\begin{center}
\begin{tabular}{|c|c|c|c|c|c|c|c|c|c|c|c|c|c|c|c|c|c|c|c|}
\hline 
$h_{x}$& 9& 1& 9& 5& 7&10& 5&14&11&14& 9& 9& 5& 5&16&10&16& 1&17 \\
\hline
$h_{y}$&11&12&14&15&15&13&15&15&16&16&17&16&17&19&17&19&20&18&21 \\
\hline
$h_{z}$&13&14&16&16&16&17&17&17&17&17&19&20&20&20&21&21&21&22&22 \\
\hline \hline
$w$ &-3& 3&-3&-3& 3&-3& 3&-3& 3& 3&-1&-2&-1& 1&-2& 5& 2&-2& 2 \\
\hline 
\end{tabular}
\end{center}
\caption{Structures and weights $w$ of the lattices used \label{lattices}}
\end{table}
and obtain the free energy coefficients up to 24th order as given in
table \ref{free_energy}.
\begin{table}
\begin{center}
\begin{tabular}{c|c} 
order k & free energy $f_{n}$ \\ \hline
0 & 0\\
2 & 0\\
4 & 3\\
6 & 22\\
8 & 375/2\\
10 & 1980\\
12 & 24044\\
14 & 319170\\
16 & 18059031/4\\
18 & 201010408/3\\
20 & 5162283633/5\\
22 & 16397040750\\
24 & 266958797382
\end{tabular}
\end{center}
\caption{Free energy up to 24th order \label{free_energy}}
\end{table}
In order to eliminate the contribution from (unphysical) loops with
an odd number of links in any direction, we use the cancellation
technique of ref. \cite{bhanot_series}. This  amounts to inserting
additional signature factors into eq. \ref{link_factors} for each
of the three link-factors
\begin{equation}
( 1 + s_0 s_{h_i} t ) \to \sigma_i ( 1 + s_0 s_{h_i} t ) 
 \mbox{   ,   } i=x,y,z
\end{equation}
with $\{\sigma_x,\sigma_y,\sigma_z\}=\{\pm,\pm,\pm\}$. By performing
8 separate runs corresponding to all possible values of $\vec{\sigma}$
and adding the results, one achieves a complete elimination of the
unwanted loops.
Possible contributions of higher-order finite-size-loops are at least 
of order 25 for this set of lattices. 
Since we use open boundary conditions, the coefficients $p$ are invariant
under the global transformation $s_i \to -s_i$. This Z(2) symmetry
enables us to reduce memory requirements by a factor of two.
Unlike refs. \cite{bhanot_ising,bhanot_series,enting} we use multiple-word 
arithmetic to account for the size of the coefficients. This implementation
needs about 100\% more memory but leads to a doubling in performance. 
Since the number of words can be adjusted separately for every order, the 
computational effort can be reduced accordingly. On the 32 node CM-5
the total time for all computations was about 50 hours. 

Compared to the finite-lattice approach of Enting and Guttmann 
\cite{enting}, our method appears to require more CPU-time since we
need to cancel unphysical loops. It should be noted, however, that 
helical lattices are very naturally implemented in data parallel software
environments and thus lead to better performance. In the usual finite
lattice method \cite{enting}, the HT expansion can only be extended in
fairly coarse steps, using lattices with $(4 \times 5)$ cross-section for 
22nd order and $(5 \times 5)$ cross-section for 26th order, respectively.
For this reason, a 24th order computation would not have been 
feasible using that method with  our computer resources.

\section{CRITICAL EXPONENT}
The specific heat is defined as
\begin{equation}
c|_{h=0}=\beta^2 \frac{\partial^2}{\partial \beta^2} \log{Z}
        =\sum_k c_k t^{2 k}
\end{equation}
and is expected to behave near $T_C$ as
\begin{equation}
c|_{h=0} = A(T) |T-T_{C}|^{-\alpha} \left[ 
          1 + B(T) |T-T_{C}|^{\theta} + \ldots \right]
          \mbox{  ,}
\end{equation}
with $A$ and $B$ being analytic near $T_{C}$ \cite{guttmann,sykes}.
We analyse the series using unbiased and biased inhomogeneous differential 
Pad\'{e}-approximants (IDPs) \cite{fisherbaker} as well as ratio-tests. 

\subsection{Pad\'{e}-Analysis}
\begin{figure}
\caption{Critical exponent $\alpha$ as a function of $t_C^2$.}
\label{idpade_24}
\epsfbox{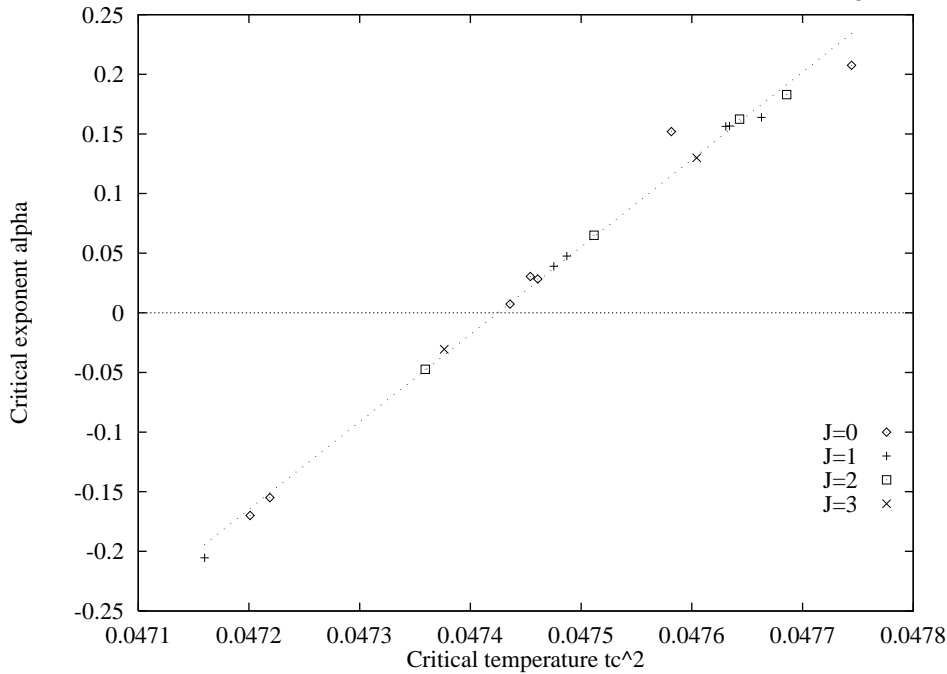}
\end{figure}
In figure \ref{idpade_24} we plot $\alpha$ against $t_{c}^2$ for 
each IDP-Approximant [J/L;M]. Fitting the linear dependence of 
$\alpha$ on $t_{C}^2$ \cite{guttmann}, we find 
\begin{equation}
\alpha=0.102 \pm 0.008
\end{equation}
at the value $t_{C}=0.218092$ as obtained in Monte-Carlo 
simulations \cite{gupta}. A direct, biased-IDP analysis was also 
performed. We obtained $\alpha=0.109 \pm 0.016$. 

IDPs can also be 
used to predict the most significant digits of the next term in the 
specific heat series \cite{enting}. The estimate of the 24th order 
term as obtained in ref. \cite{enting} agrees perfectly with our 
exact result. Using the same method we can estimate the 26th order term 
in the expansion to be 
\begin{equation}
f_{26}=443762(4) \times 10^7 \mbox{  ,} \label{extrapol}
\end{equation}
where the  errors quoted are two standard deviations.

\subsection{Ratio-Test}
The main problem in the determination of critical exponents in the
low-temperature case is the presence of unphysical singularities
nearer to the origin than the physical one. Since the expansion
coefficients $c_n$ are dominated by these unphysical singularities,
ratio-methods cannot be applied. 

\begin{figure}
\epsfbox{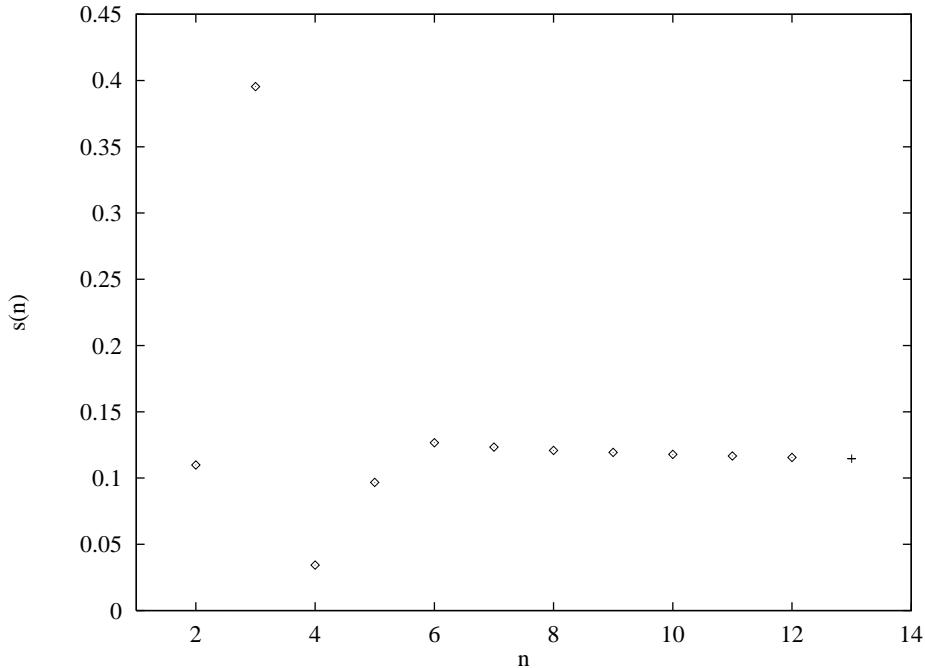}
\caption{Plot of the sequence $s_n$ against $n$. The error of $s_{13}$, 
obtained from the ID-Pad\'{e} extrapolation, is too small to be visible.
\label{s_sequence}}
\end{figure}

\begin{figure}
\epsfbox{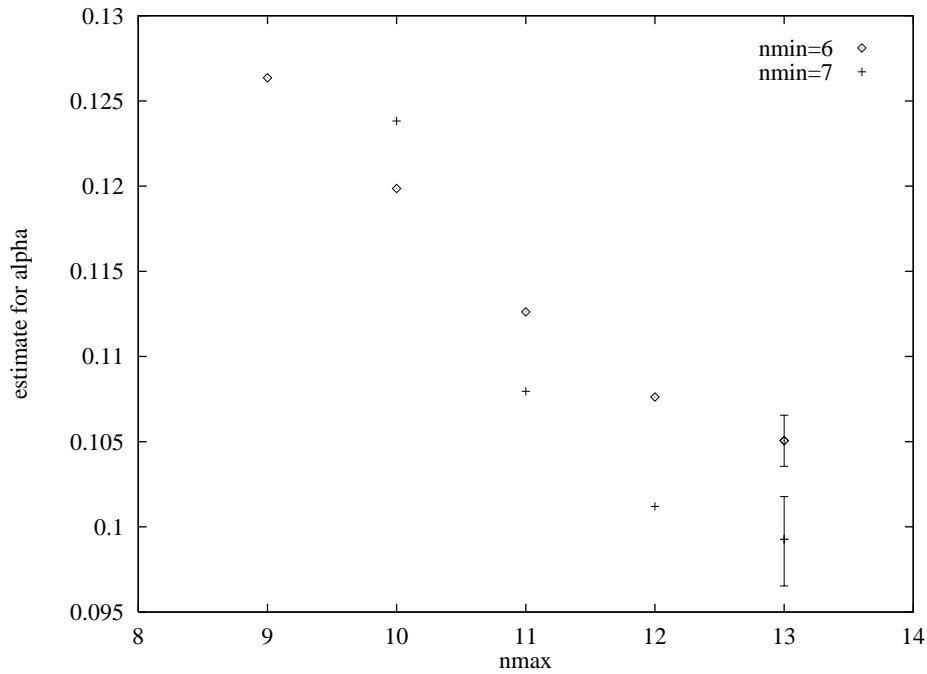}
\caption{Estimates of alpha using a 3-parameter-fit. Each point
represents the results of a fit to the set of values $\{s_{n_{min}}
,\ldots,s_{n_{max}}\}$. The error bars of the rightmost values 
represent the uncertainty of the extrapolated 13th term.
\label{alpha}}
\end{figure}

In the HT-expansion, the physical singularity dominates the asymptotic 
behaviour, so that the ratio $r_n=c_n/c_{n-1}$ of successive 
coefficients of the series is expected to behave as \cite{guttmann}
\begin{equation}
r_{n}=\frac{1}{t_C^2} \left( 1 + \frac{\alpha-1}{n} + 
       \frac{c}{n^{1 +\theta}} + \frac{d}{n^{1 + 2 \theta}} +
       O\left(\frac{1}{n^{1+ 3 \theta}}\right) 
       \right) \mbox{  .} 
\end{equation}
Assuming that the correction-to-scaling exponent $\theta$ is close
to 0.5 \cite{gupta,fisher}, the following sequence $s_{n}$ is expected 
to converge towards $\alpha$ like
\begin{equation}
s_{n}:=\left( t_C^2 r_n -1 \right) n +1 
      =\alpha + \frac{c}{n^{1/2}} + \frac{d}{n} + O\left(\frac{1}{n^{3/2}}
       \right) \mbox{  .}
       \label{fit_ansatz}
\end{equation}
A plot of this sequence against $n$ is shown in figure \ref{s_sequence}.
Obviously the first four values are dominated by higher order corrections.
To obtain estimates for $\alpha$ we therefore use only the values 
$\{s_6,\ldots,s_{13}\}$. A 3-parameter least-square-fit using the ansatz
of eq. (\ref{fit_ansatz}) yields the values shown as diamonds in figure 
\ref{alpha}.
The value of $\alpha=0.113$ obtained by the fit to the points
$\{s_6,\ldots,s_{11}\}$ is in perfect agreement with the result of ref. 
\cite{enting}. Their estimate of $\alpha=0.110$ using the
extrapolated term $s_{12}$ appears to be slightly above our
value of $\alpha=0.108$ using the exact term. Including our 
value for $s_{13}$ of the ID-Pad\'{e} extrapolation eq. (\ref{extrapol})
we obtain $\alpha=0.105(2)$. The error represents the uncertainty
of the extrapolation. 
However, from fig. \ref{alpha} it it quite suggestive 
that the $\alpha$-values might converge to a value below $0.105$. 

To get an estimate of the uncertainties of our results, we investigate 
the stability of the fits. For this purpose, we repeat the analysis
after eliminating the point 
$s_6$ from the data. As a result we obtain sizeable changes
for $\alpha$. The new data are shown as crosses in fig. \ref{alpha}.

\begin{figure}
\epsfbox{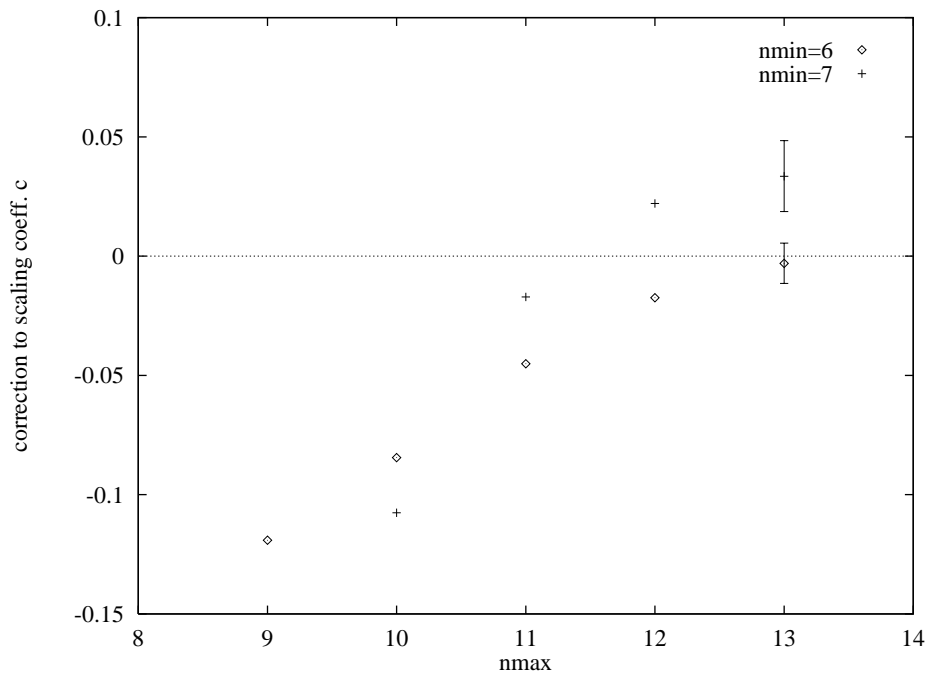}
\caption{Estimates of c using a 3-parameter-fit. Each point
represents the results of a fit to the set of values $\{s_{n_{min}},\ldots,
s_{n_{max}}\}$. The error bars of the rightmost values represent the
uncertainty of the extrapolated 13th term.\label{c_coeff}}
\end{figure}

\begin{figure}  
\epsfbox{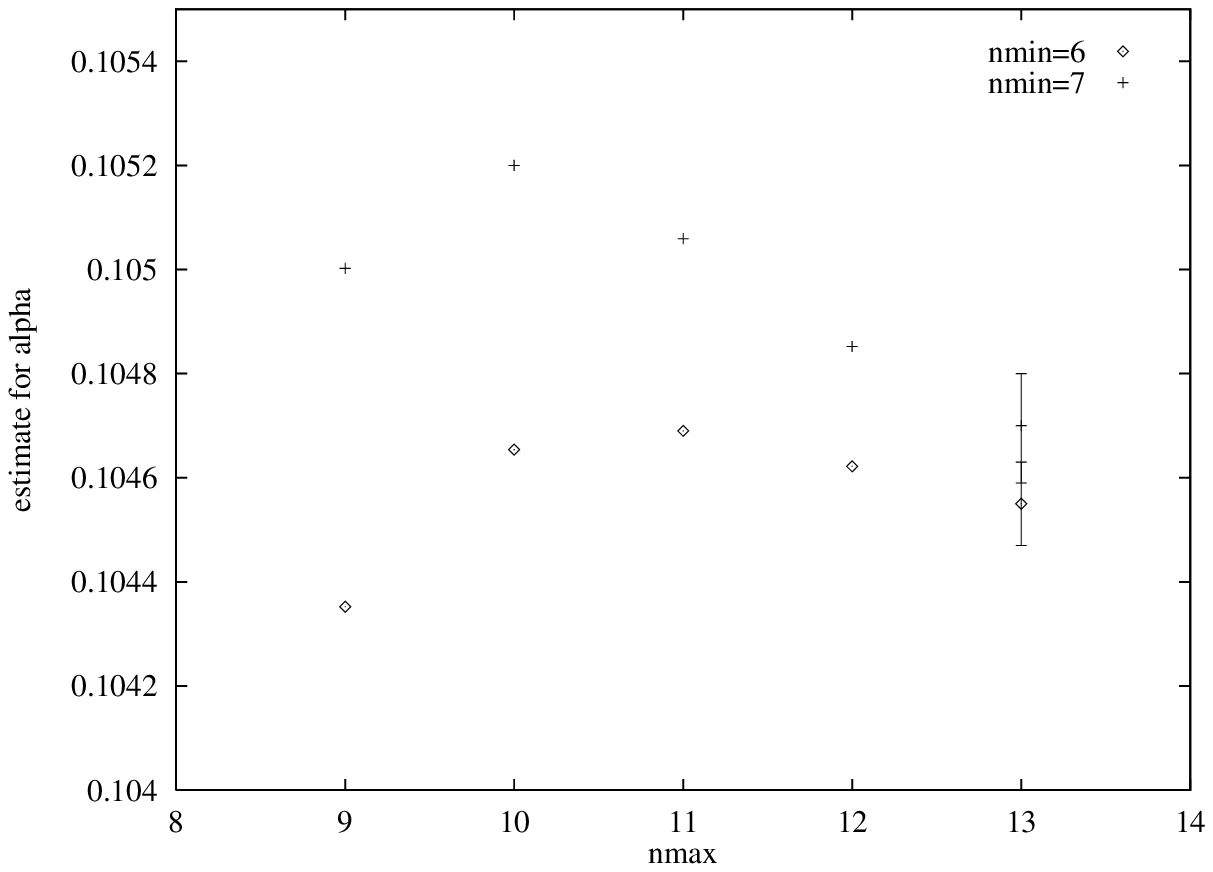}
\caption{Estimates of $\alpha$ using a 2-parameter-fit with
$c=0$. Each point represents the results of a fit to the set of values 
$\{s_{n_{min}},\ldots,s_{n_{max}}\}$. The error bars of the rightmost 
values represent the uncertainty of the extrapolated 13th term.
\label{alpha_lin}}
\end{figure}

In figure \ref{c_coeff} we present the results for the 
first correction-to-scaling coefficient $c$ from our 3-parameter fits.
In contrast to ref. \cite{enting}, our values suggest that $c$ changes sign 
with increasing $n_{max}$. Because of the sensitivity of the fits to 
the number of terms we keep, 
it is difficult to determine the value of $c$ very precisely. Our best 
estimate is $c=0.01(4)$. Since $c$ vanishes within error, it seems 
reasonable to also try a 2-parameter-ansatz with $c=0$ to fit the data. The 
results of these fits are shown in figure \ref{alpha_lin}.
We now find that the fits are much more stable and the 
$\alpha$ estimates show much more of a convergence to their asymptotic values.
The best value (from the largest $n_{max}$) is $\alpha=0.1045(3)$.
This value supports the impression of the 3-parameter fits, which suggested 
that $\alpha$ was slightly below $0.105$. Taking into account the fact that
neglecting $c$ causes a systematic error, our final estimate for the 
critical exponent is,
\begin{equation}
\alpha=0.104(4) \mbox{  .} \label{alpha_value}
\end{equation}

\section{DISCUSSION AND OUTLOOK}
The crucial element in the estimate of the error 
in $\alpha$ ( eq.~\ref{alpha_value} ) is our neglect of the 
correction-to-scaling coefficient $c$. The resulting systematic error 
is rather large. From fig. \ref{c_coeff} one might speculate that the
estimates for $c$ begin to exhibit asymptotic behaviour at the 26th
order. Therefore an exact calculation of the 26-th term of the expansion 
might reduce the uncertainty of $c$ significantly. If the magnitude of 
$c$ turns out to be really negligible, one could adopt the errors of the 
linear fits, and $\alpha$ would be obtained accurate to the fourth 
significant digit.

\section{ACKNOWLEDGEMENTS}
This work was partly funded under Contracts No. DE-AC02-76CH00016 and
DE-FG02-90ER40542 of the U.S.~Department of Energy.  Accordingly, the
U.S.~Government retains a nonexclusive, royalty-free license to publish or
reproduce the published form of this contribution, or allow others to do
so, for U.S. Government purposes. The work of GB was also partly supported
by a grant from the Ambrose Monell Foundation. UG and KS are grateful
to Deutsche Forschungsgemeinschaft for its support to the Wuppertal
CM-Project.

\end{document}